# Clarification as to why alcoholic beverages have the ability to induce superconductivity in $Fe_{1+d}Te_{1-x}S_x$

K Deguchi[1,2,3], D Sato[4], M Sugimoto[4], H Hara[1], Y Kawasaki[1,2,3], S Demura[1,2,3], T Watanabe[1,2], S J Denholme[1], H Okazaki[1,3], T Ozaki[1,3], T Yamaguchi[1,3], H Takeya[1,3], T Soga[4], M Tomita[4] and Y Takano[1,2,3]

[1] National Institute for Materials Science, 1-2-1, Sengen, Tsukuba, 305-0047, Japan

[2] Graduate School of Pure and Applied Sciences, University of Tsukuba, 1-1-1 Tennodai, Tsukuba, 305-8571, Japan

[3] JST-EU-Japan, 1-2-1, Sengen, Tsukuba, 305-0047, Japan

[4] Institute for Advanced Biosciences, Keio University, 246-2, Mizukami, Kakuganji, Tsuruoka, Yamagata 997-0052, Japan

E-mail : DEGUCHI.Keita@nims.go.jp

Abstract

To elucidate the mechanism as to why alcoholic beverages can induce superconductivity in $Fe_{1+d}Te_{1-x}S_x$ samples, we performed component analysis and found that weak acid such as organic acid has the ability to induce superconductivity. Inductively-coupled plasma spectroscopy was performed on weak acid solutions post annealing. We found that the mechanism of inducement of superconductivity in $Fe_{1+d}Te_{1-x}S_x$ is the deintercalation of excess Fe from the interlayer sites.

Keywords: Superconductivity, Fe-based superconductor, FeTe, Alcoholic beverage annealing, Deintercalation

# 1. Introduction

Since the discovery of superconductivity in LaFeAsO$_{1-x}$F$_x$, research in Fe-based superconductivity has been actively performed and several types of Fe-based superconductors have been discovered [1-4]. Among them, the 11 system, whose parent compounds are FeSe and FeTe. These have advantages for understanding the mechanism of Fe-based superconductivity because they are binary compounds and have the simplest crystal structure. FeSe shows superconductivity with a transition temperature $T_c$ of 8.5 K. In contrast to FeSe, isostructural FeTe does not show superconductivity and exhibits an antiferromagnetic transition around 70 K. The S substitution for the Te site in FeTe suppresses the antiferromagnetism and induces superconductivity [5].

Fe$_{1+d}$Te$_{0.8}$S$_{0.2}$ synthesized using a melting method showed a shielding volume fraction of about 20 %, while it contained minor impurity phases. In contrast, almost pure Fe$_{1+d}$Te$_{0.8}$S$_{0.2}$ is obtained using a solid-state reaction. On the other hand, the as-grown sample does not show superconductivity, although the antiferromagnetic ordering seems to be suppressed. Superconductivity in the solid-state reacted sample however can be induced by air exposure, water immersion, and oxygen annealing [6-8]. Recently, it has been revealed that oxygen annealing is also effective for superconductivity in Fe$_{1+d}$Te$_{1-x}$Se$_x$ and it is proposed that oxygen suppresses the magnetic moment of the excess Fe [9].

Furthermore we reported that hot alcoholic beverages were more effective in inducing superconductivity in $Fe_{1+d}Te_{0.8}S_{0.2}$ than water [10]. The shielding volume fractions of the samples heated in red wine, white wine, beer, Japanese sake (rice wine), whisky, and shochu (distilled spirit) are 62.4, 46.8, 37.8, 35.8, 34.4, and 23.1 %, respectively. These values are clearly higher than immersing the sample in pure water and ethanol solutions. In a previous report, we concluded that alcoholic beverages can induce superconductivity, although the exact mechanism of how they act to enhance the superconductivity in $Fe_{1+d}Te_{1-x}S_x$ remains unsolved. To reveal the mechanism, we utilized a technology of metabolomic analysis, a simultaneous profiling of hundreds of small molecules, to analyze the ingredient in the alcoholic beverages and other solutions heated with the samples. In this study, we investigated the key components in alcoholic beverages and the systematic mechanism to induce superconductivity in $Fe_{1+d}Te_{0.8}S_{0.2}$.

## 2. Experimental

### 2-1. Sample preparation

Polycrystalline samples of $Fe_{1+d}Te_{0.8}S_{0.2}$ were prepared using a solid-state reaction method. Powders of Fe and TeS, and grains of Te with a nominal composition of $FeTe_{0.8}S_{0.2}$ were put into a quartz tube. The quartz tube was then evacuated by a rotary pump and sealed. After being heated at 600 °C for 10 hours, the obtained mixture was ground, pelletized, and put into a quartz tube. The quartz tube was pumped, sealed, and heated again at 600 °C for 10 hours. We prepared 5 glass bottles filled with different liquids: ultrapure water, red wine (Bon Marche, Mercian Corporation), and aqueous solutions of malic acid, citric acid, or $\beta$-alanine. Here these acids were dissolved in ultrapure water. The concentrations of malic acid, citric acid, and $\beta$-alanine in the solutions were 0.16, 0.12, and 0.004 g/L, respectively, concentrations equal to those found in the red wine. The sintered $Fe_{1+d}Te_{0.8}S_{0.2}$ pellet was put into each liquid and heated at 70 °C for 24 hours.

Powder x-ray diffraction patterns were measured using the $2\theta/\theta$ method with the Cu $K\alpha$ radiation. The temperature dependence of susceptibility was measured using a SQUID magnetometer down to 2 K under a magnetic field of 10 Oe. The shielding volume fraction was estimated from the lowest-temperature value of the magnetic susceptibility after zero-field cooling.

**2-2. Component analysis**

Metabolomic analysis of the alcoholic beverages was performed using a capillary electrophoresis time-of-flight mass spectrometer (CE-TOFMS) with slight modifications [11, 12]. The alcoholic beverages used in this study were red wine (Bon Marche), white wine (Bon Marche, Mercian Corporation), beer (Asahi Super Dry, Asahi Breweries, Ltd.), Japanese sake (Hitorimusume, Yamanaka shuzo Co., Ltd.), shochu (The Season of Fruit Liqueur, TAKARA Shuzo Co., Ltd.) and whisky (The Yamazaki Single Malt Whisky, Suntory Holdings Limited). These alcoholic beverages were centrifuged at 5,800 × $g$ for 15 min (4 °C) to remove the sediments. The supernatants were filtrated with Ultrafree-MWCO 5,000 centrifugal filter unit (Millipore) to eliminate large molecules. The filtrated solutions were diluted ten-fold with water including the internal standards (methionine sulfone, 3-aminopyrrodine, D-camphor-10-sulfonic acid, and trimesic acid, 200μM each), and applied to CE-TOFMS. The Fe concentration of the solution after annealing with the sample was analyzed using inductively-coupled plasma (ICP) spectroscopy.

**3. Results and discussions**

CE-TOFMS successfully quantified several hundreds of charged metabolites, such as amino acids, organic acids, nucleotides, and dipeptides. We calculated correlation coefficients between concentration of a metabolite and the shielding volume fractions obtained from the samples annealed in alcoholic beverages. The value of correlation coefficient took on a wide range of 0.947 to -0.631. Among the metabolites, we focused on malic acid, citric acid, and $\beta$-alanine. Malic acid had a quite high value of 0.907 and is present at high concentration in alcoholic beverages compared to other metabolites with a higher value. Citric acid was about the same concentration as malic acid but with a slightly reduced value of 0.675. $\beta$-alanine showed the highest value of 0.947. Figure 1 shows the shielding volume fractions of the samples with various alcoholic beverages in ref. 10 as functions of the concentrations of three acids: (a) malic, (b) citric, and (c) $\beta$-alanine. The shielding volume fractions of the samples annealed in the alcoholic beverages was increased in proportion to the concentrations of these compounds.

Figure 2 shows the x-ray diffraction patterns for the as-grown $Fe_{1+d}Te_{0.8}S_{0.2}$ sample and the samples annealed in the red wine, malic acid, citric acid, $\beta$-alanine, and water. The calculated FeTe peaks, taken from PDF#01-089-4077, were also plotted at the bottom. There are no significant differences among all the patterns, indicating that heating the samples in the solutions does not decompose the crystal structure.

Figure 3 shows the magnetic susceptibility versus temperature for the as-grown sample and the samples annealed in various liquids at 70 °C for 24 hours. All the samples annealed in the solutions show superconductivity, whereas no superconducting signal is observed in the as-grown sample. We estimate the shielding volume fraction of the samples annealed in the red wine, malic acid, citric acid, $\beta$-alanine, and water to be 56.0, 35.6, 34.8, 25.5, and 17.6 %, respectively. The obtained shielding volume fractions are summarized in Fig. 4. We found that the samples annealed in solutions of malic acid, citric acid, and $\beta$-alanine have a larger shielding volume fraction compared with the sample annealed in water, and the volume fraction increases with decreasing pH.

In order to understand the role of the malic acid, citric acid, and $\beta$-alanine, we performed ICP analysis for these solutions after annealing with samples to estimate the Fe concentration. Figure 5 shows the concentration of Fe dissolved in solutions as a function of pH. The average concentration of Fe in red wine, malic acid, citric acid, $\beta$-alanine, and water are 85.8 ± 0.91 (standard deviation), 58.0 (± 0.17), 37.2 (± 0.49), 4.5 (± 0.17), and 4.3 (± 0.02) ppm respectively, which corresponds to 2.81, 1.90, 1.22, 0.15, and 0.13 % of Fe in the sample. The value of red wine after annealing without a sample is also plotted in Figure 5, indicating that red wine itself contained only a small quantity of Fe. In comparison with figure 4, it is obvious that the concentration of Fe in solutions, that is the decrement of Fe from the sample, is related to the shielding volume fraction.

It is known experimentally that FeTe always has some amount of excess Fe at the interlayer sites [13]. A previous report indicated that excess Fe is the cause of the bicollinear antiferromagnetic order and is not in favor of superconductivity [14]. Therefore, reducing the excess Fe is required to achieve superconductivity. As mentioned above, the shielding volume fractions of $Fe_{1+d}Te_xS_x$ were increased with reducing the Fe in the sample. Furthermore, as shown in figure 2, the crystal structure of the samples was not decomposed by annealing. These results suggest that part of the excess Fe was deintercalated from the interlayer sites. Therefore, weak acid annealing suppresses the antiferromagnetic correlation by deintercalating the excess Fe and, hence superconductivity is achieved.

To summarize, we investigated the key components in alcoholic beverages and clarified the mechanism to induce superconductivity in $Fe_{1+d}Te_{1-x}S_x$. The shielding volume fractions of the samples annealed in the alcoholic beverages were increased in proportion to the concentrations of malic acid, citric acid, and $\beta$-alanine. From ICP analysis, we found that the shielding volume fractions of $Fe_{1+d}Te_{1-x}S_x$ were increased with reducing the excess Fe in the sample. This means that weak acid has the ability to deintercalate the excess Fe in the sample. It is concluded that the inducement of superconductivity in $Fe_{1+d}Te_{0.8}S_{0.2}$ is achieved by soft chemical reaction using weak acid such as organic acid.

**Acknowledgement**

This work was partly supported by a Grant-in-Aid for Scientific Research (KAKENHI) and by research funds from the Yamagata Prefectural Government and the city of Tsuruoka.

**Figure captions**

Figure 1. The shielding volume fraction of $Fe_{1+d}Te_{0.8}S_{0.2}$ samples annealed in various alcoholic beverages as a function of the concentration of (a) malic acid, (b) citric acid, (c) $\beta$-alanine. The volume fractions presented in this figures were obtained in the previous study (ref. 10).

Figure 2. The powder x-ray diffraction patterns of an as-grown $Fe_{1+d}Te_{0.8}S_{0.2}$ sample, samples annealed in various liquids. The sharp lines on the bottom of the figure indicate $2\theta$ and intensity expected from a calculation.

Figure 3. The temperature dependence of magnetic susceptibility for the as-grown $Fe_{1+d}Te_{0.8}S_{0.2}$ sample and the samples annealed in various solutions. The $\beta$-alanine, citric acid, and malic acid were dissolved in water.

Figure 4. The shielding volume fraction estimated from the lowest-temperature value of magnetic susceptibility for the samples annealed in various liquids as a function of pH.

Figure 5. The pH dependence of Fe concentration dissolved in solutions after annealing with the sample.

Figure 1a

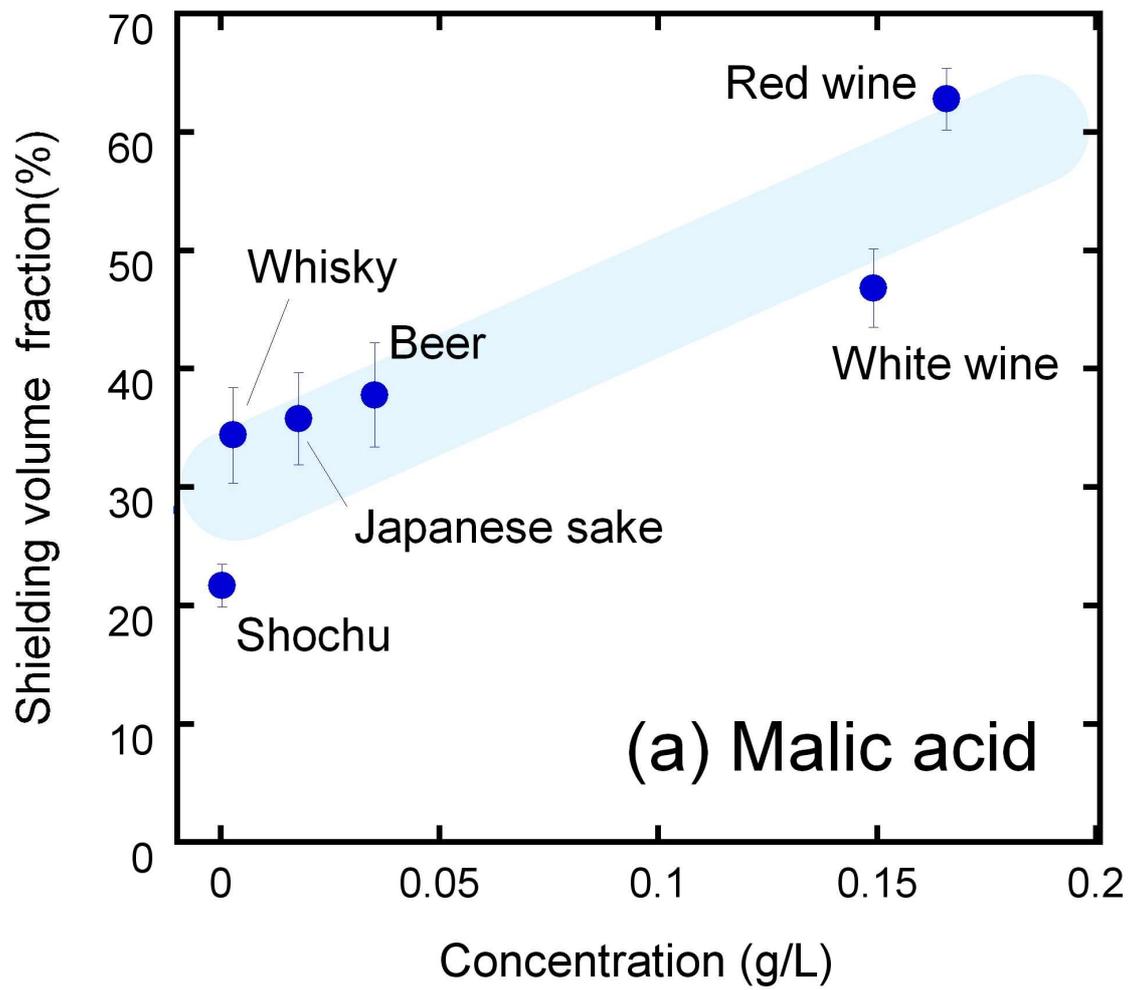

Figure 1b

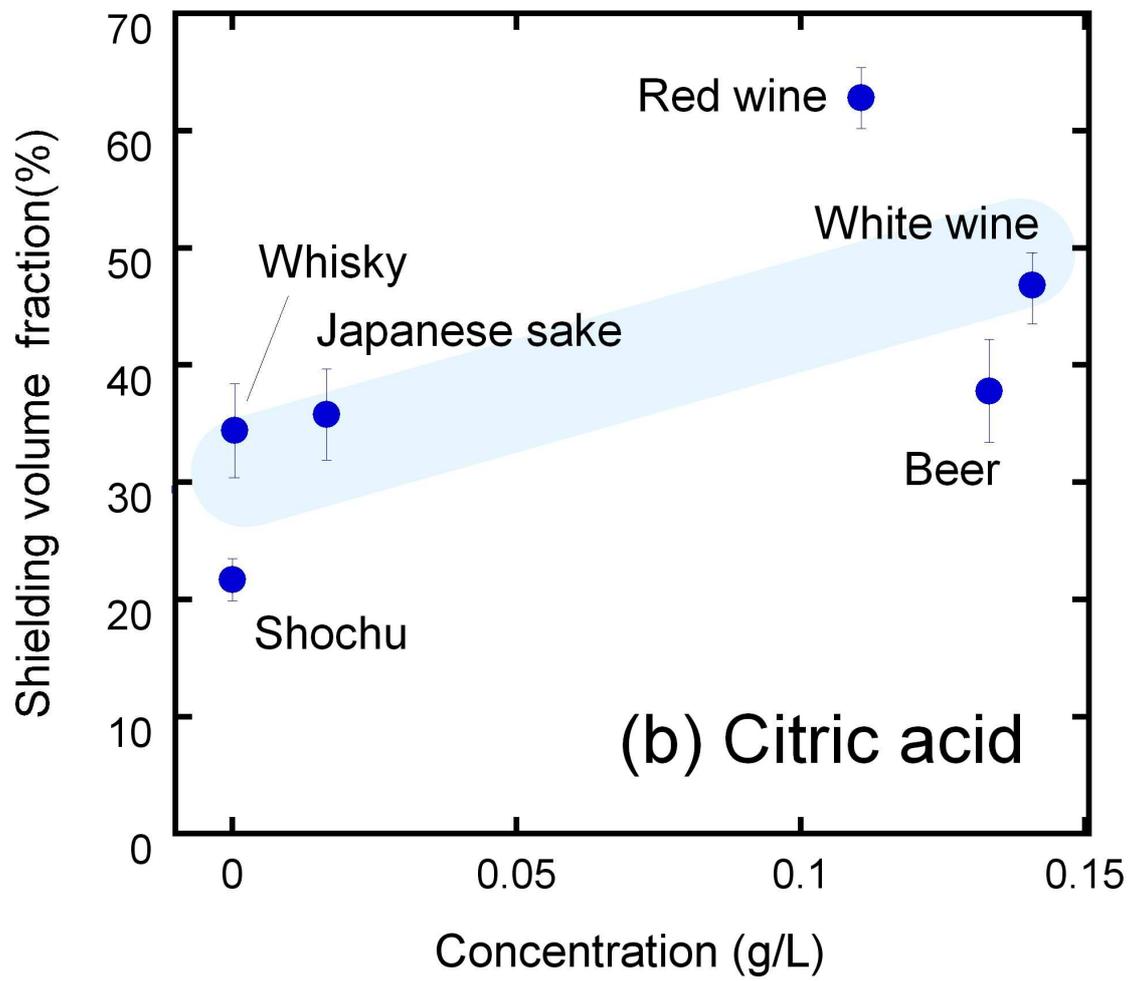

Figure 1c

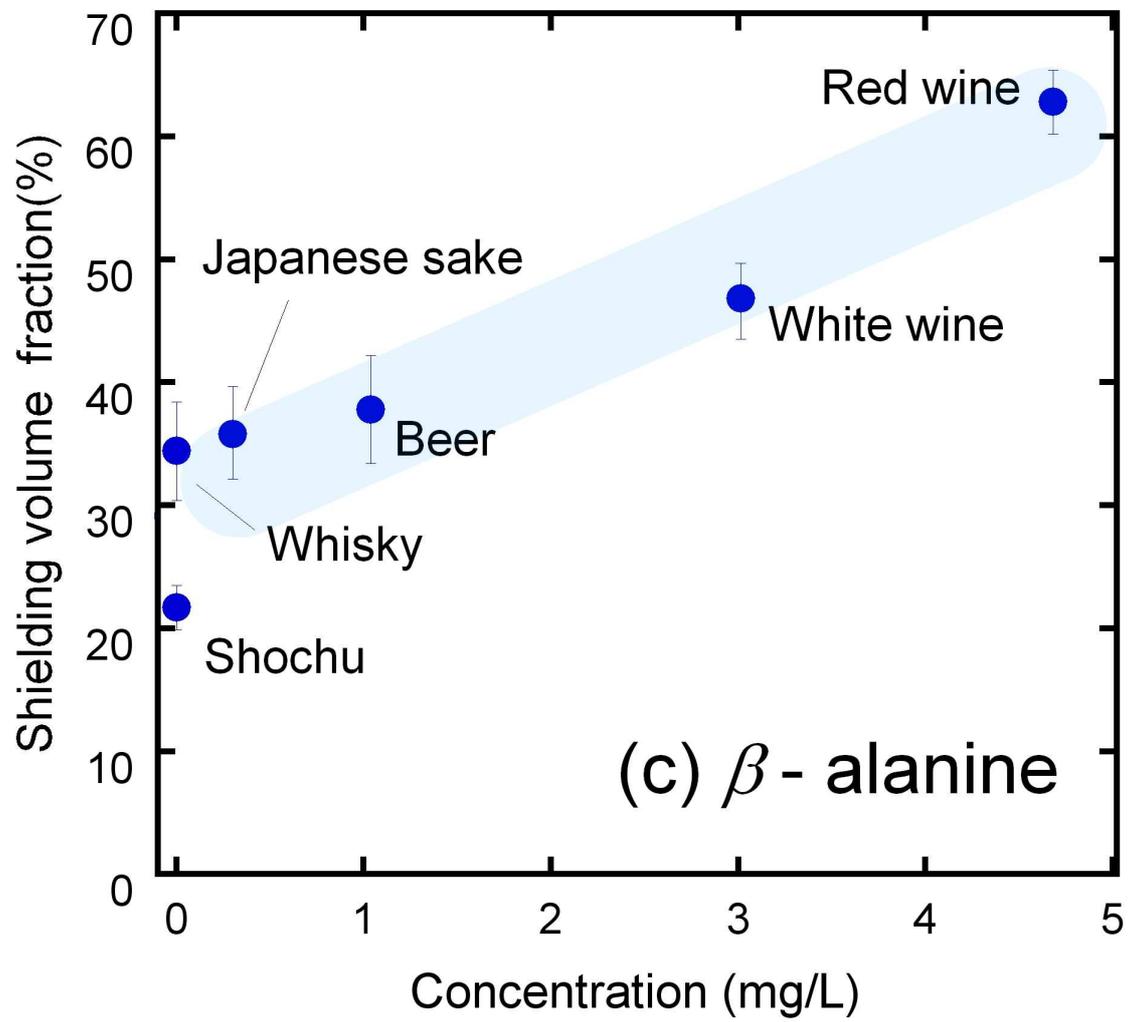

Figure 2

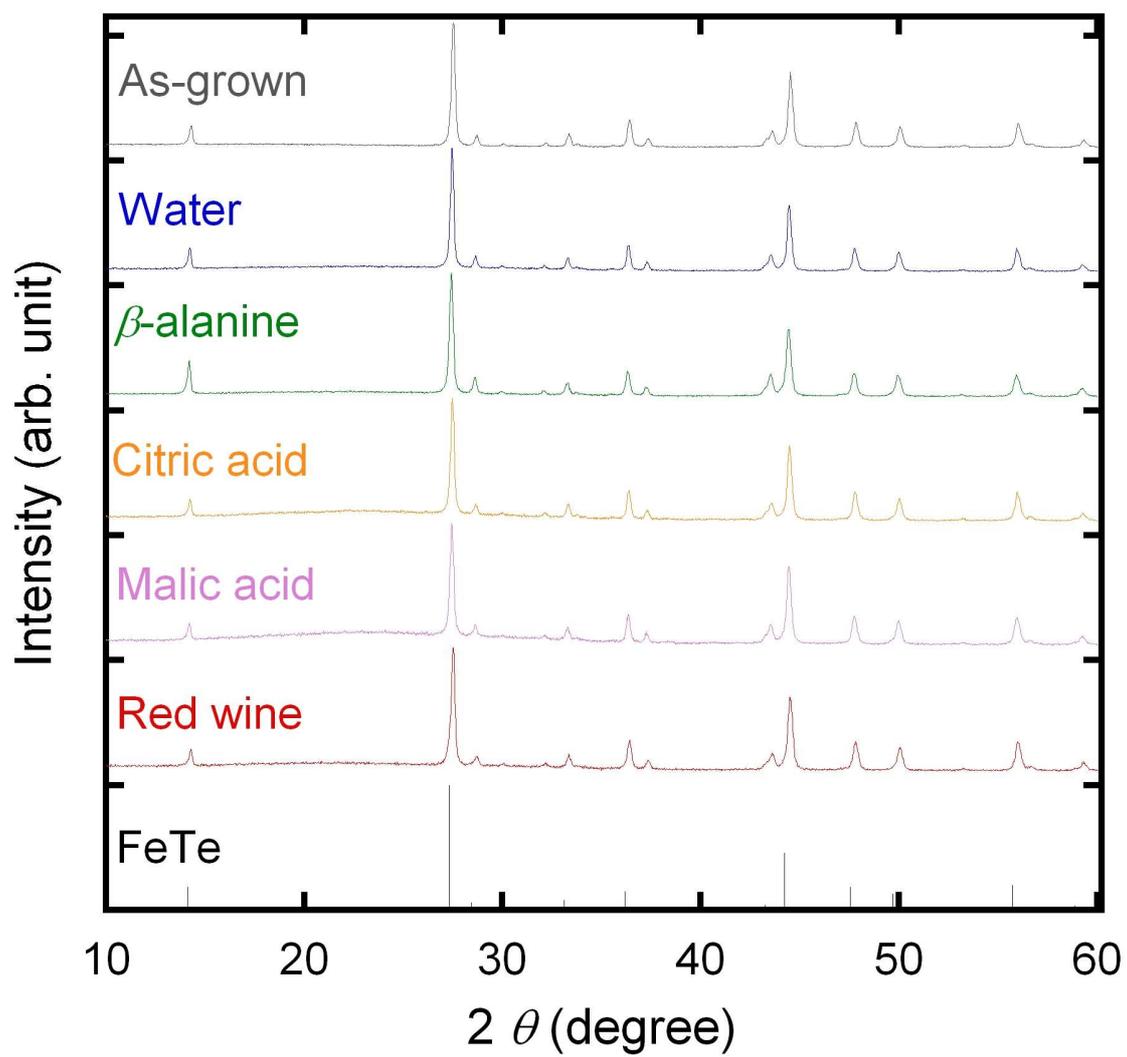

Figure 3

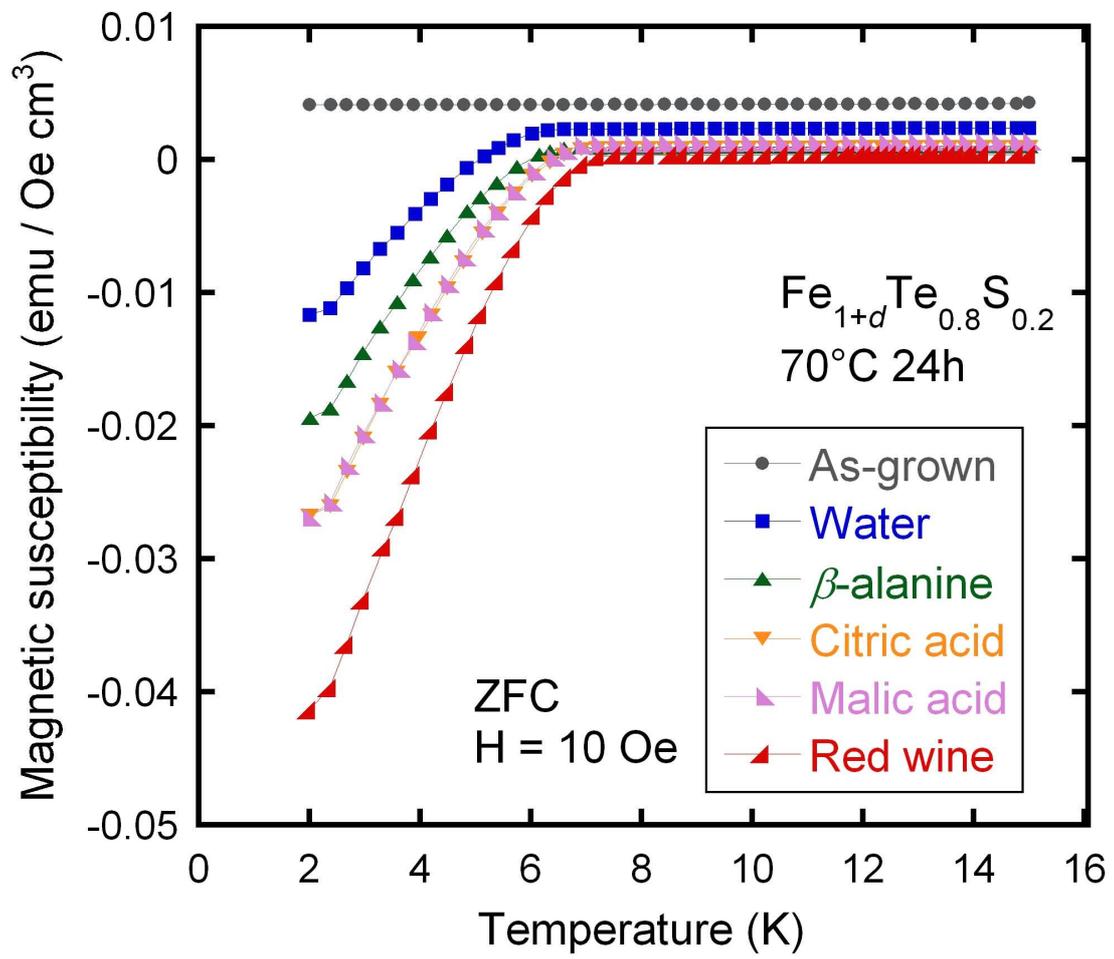

Figure 4

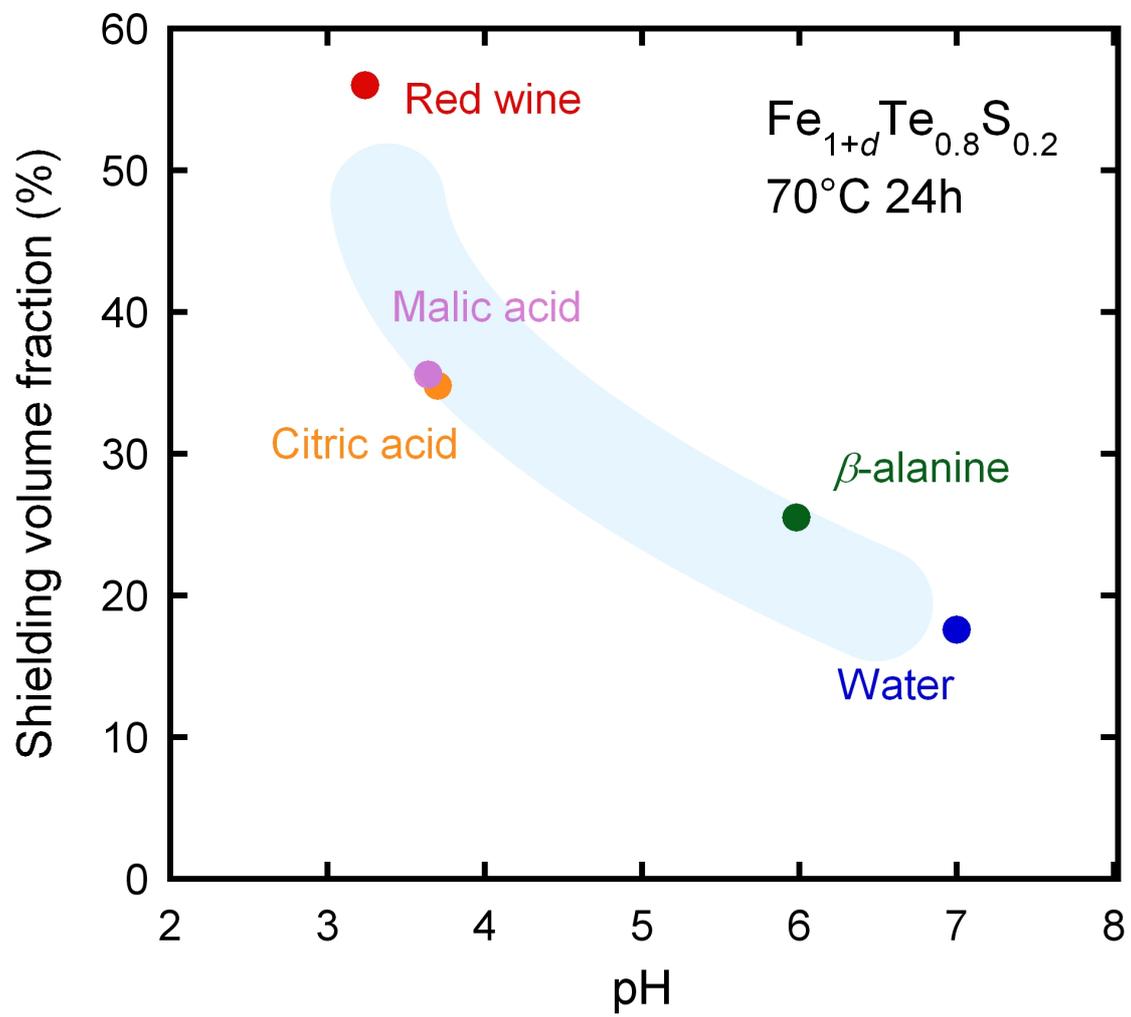

Figure 5

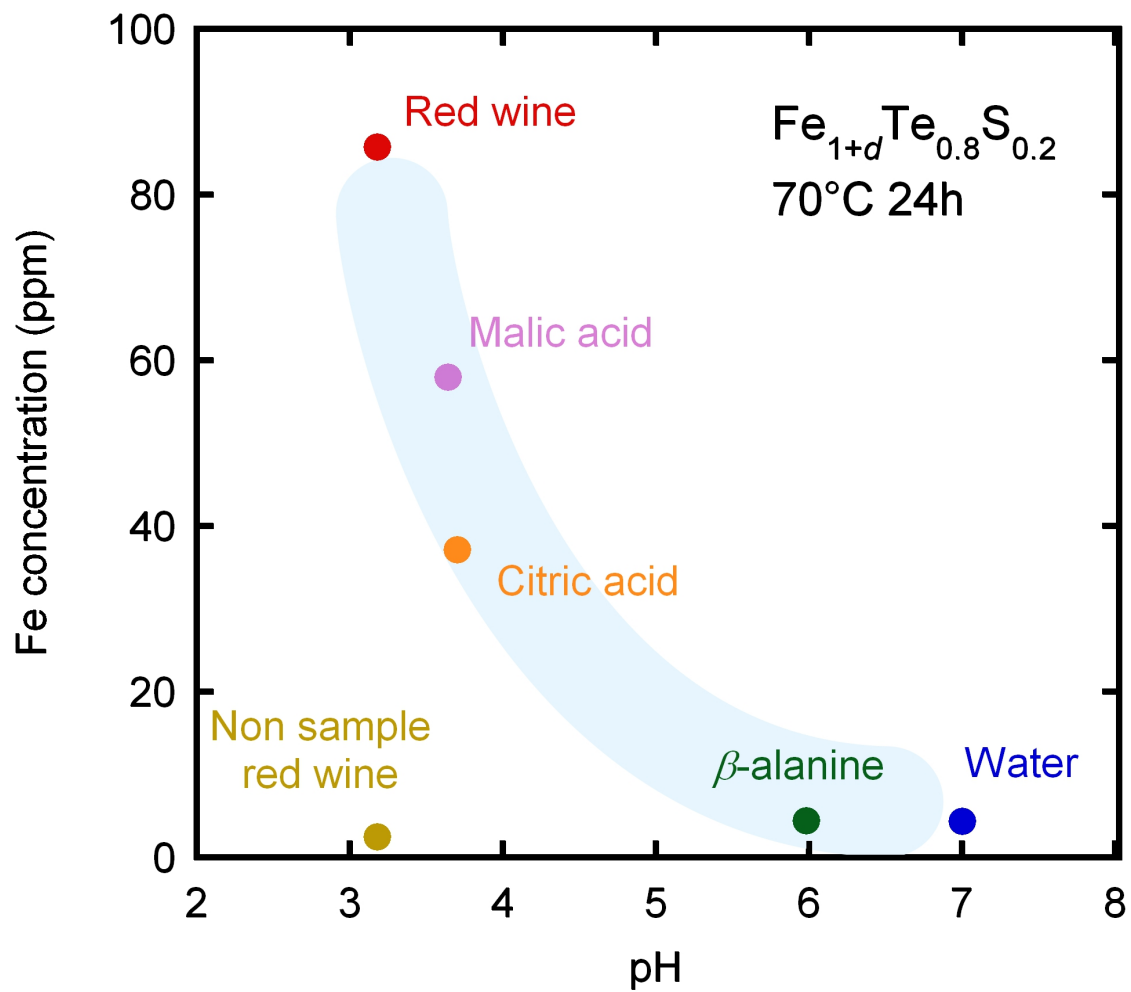